\documentclass[aps,prx,twocolumn,superscriptaddress]{revtex4-2}
\usepackage[utf8]{inputenc}
\usepackage{xcolor}
\usepackage{graphicx}
\usepackage{float}
\usepackage{latexsym}
\usepackage{amsmath}
\usepackage{amsfonts}
\usepackage{amssymb}
\usepackage{bm}
\usepackage{calc}
\usepackage{mathtools}
\usepackage{amsbsy}
\usepackage{physics}
\usepackage{enumitem}
\usepackage{media9}
\usepackage{hyperref}
\usepackage{multirow}
\graphicspath{\figures}

\begin{document}

\title{ Force-induced Elastic Softening and Conformational Transitions in a Polyampholyte
chain }

\author{Rakesh Palariya}%
\email{rakesh20@iiserb.ac.in}
\affiliation{Department of Physics, Indian Institute of Science Education and Research, \\ Bhopal 462 066, Madhya Pradesh, India}

\author{Sunil P Singh}
 \email{spsingh@iiserb.ac.in}
\affiliation{Department of Physics, Indian Institute of Science Education and Research, \\ Bhopal 462 066, Madhya Pradesh, India}

\begin{abstract}
{
The mechanical response of intrinsically disordered proteins (IDPs) and polyampholyte (PA) chains is vital for understanding their biological functions and designing functional materials. We investigate the force-extension behavior of a PA chain with distinct charge sequences using molecular dynamics simulations and a theoretical approach based on the generalized random-phase approximation (GRPA). A diblock PA chain under extensional force undergoes a continuous coil-to-stretch transition at weak electrostatic coupling, which sharpens into a globule-coil-like transition at stronger coupling. The GRPA theory quantitatively captures these behaviors, including the sharp conformational transition and its dependence on electrostatic strength. Simulations reveal pronounced hysteresis during the force-extension and relaxation processes. 
Additionally, the elastic modulus exhibits four regimes: an initial plateau, stress stiffening,  an exponential stress-softening behavior, and a stress stiffening regime. Using the theoretical model and structural input of the PA chain, we have demonstrated that the elastic modulus in the elastic softening regime decreases exponentially, $E\sim \exp(-\alpha_0 f/\Gamma_e)$, as a function of $f$, which aligns with the simulation results. The elastic response of the PA chain is further examined across different charge sequences, where both elastic softening and sharp transitions are absent at smaller block lengths. Finally, coarse-grained models of IDPs such as LAF-1 and DDX4 exhibit similar nonlinear elasticity, highlighting the universality of these mechanisms. Our results establish a fundamental link between electrostatic correlations, charge sequence, and nonlinear elasticity, bridging molecular interactions and macroscopic mechanics.}

\end{abstract}

\maketitle
\section{ Introduction }
{
A wide variety of biopolymers—including nucleic acids, and intrinsically disordered proteins (IDPs)—as well as synthetic macromolecular systems, contain both acidic and basic functional groups \cite{lowe2002synthesis,dobrynin2004polyampholytes,higgs1991theory}.
These systems play essential roles in numerous biological processes, such as cell signaling \cite{bondos2022intrinsically,arai2015conformational}, transcriptional regulation \cite{shang2024intrinsically,miao2025roles}, muscle elasticity \cite{tskhovrebova1997elasticity}, liquid–liquid phase separation (LLPS), and the formation of biomolecular condensates \cite{mo2022liquid,andre2020liquid,mukherjee2023liquid,brangwynne2015polymer,banani2017biomolecular,brangwynne2009germline}.
Understanding their mechanical response is crucial for elucidating the behavior of biological networks \cite{rubinstein2003polymer}, protein functionality \cite{bondos2022intrinsically,arai2015conformational,shang2024intrinsically,miao2025roles}, and the folding energy landscapes of RNA and DNA \cite{camunas2016elastic,ahlawat2021elasticity,saleh2009nonlinear,saleh2015perspective}.
Many such polymers and proteins possess folded structural domains that undergo cooperative unfolding under applied forces, resulting in complex force–extension curves (FECs) characterized by abrupt extensions and nonlinear mechanical responses \cite{andre2007forced,cardenas2009stick,rief1997reversible,dietz2004exploring,forbes2005titin}.

The theoretical frameworks for the elastic response of a polymer have led to interesting scaling relationships that describe mechanical behavior across different regimes\cite{pincus1976excluded,rubinstein2003polymer,guth1934innermolekularen,kuhn1936beziehungen,james1943theory,pincus1976excluded,pierleoni1997signature,saleh2009nonlinear,li2015modeling}. These relationships have been supported by both computational approaches and experimental observations\cite{saleh2009nonlinear,truong2023pincus,stevens2012simulations,stevens2013simulations,seol2004elastic,radhakrishnan2019force,chen2024nonlinear, andre2007forced,cardenas2009stick,rief1997reversible,dietz2004exploring,forbes2005titin}.
The single‑molecule techniques such as atomic force microscopy (AFM) and optical/magnetic tweezers have significantly transformed the study of the polymer's elastic behavior\cite{neuman2008single,mcintosh2009detailed,ahlawat2021elasticity,seol2004elastic,camunas2016elastic,innes2021flexible,giannotti2007interrogation,bao2020environment,bustamante2003ten,saleh2009nonlinear,seol2004elastic,hoffmann2012single}.  These methods enable direct measurement of force–extension relationships at the level of individual chains, thereby validating theoretical predictions, identifying distinct mechanical regimes, and highlighting how molecular heterogeneity influences bulk mechanical behavior.

The elastic response of a linear, uncharged polymer in the low-force regime is characterized by purely entropic behavior, with the force–extension curve being linear, and the elastic modulus ($E$) remaining constant. At large stretching forces, a nonlinear relationship emerges between force and extension. In the case of the freely jointed chain (FJC) model\cite{khokhlov1994statistical}, the extension approaches $<R_e> \approx R_{max} [1 - k_BT/f\ell_K]$, where $R_e$, $R_{max}$, $k_BT$, and $\ell_K$ are the chains extension, contour length, thermal energy, and Kuhn length, respectively. Alternatively, for semiflexible chains such as double-stranded DNA, a worm-like chain (WLC) model developed by Marko and Siggia  provides the force extension as  $R_e \approx R_{max} [1 - \sqrt{k_BT/2f
\ell_K]}$ \cite{pincus1976excluded,morrison2007stretching,mcintosh2009detailed,toan2010theory,saleh2015perspective,marko1995stretching,bustamante1994entropic}. 
The WLC model provides a fairly good description of the elastic response of ssRNA, ssDNA, PolyEthylene Glycol, and various other proteins across a broad range of forces\cite{seol2004elastic,linke2002pevk,li2001multiple,saleh2009nonlinear,dittmore2011single,rubinstein2003polymer}.  In these large-force limits, the elastic modulus scales as $E\sim f^2$ for the FJC and $E\sim f^{1.5}$ for the WLC model\cite{dobrynin2010chains}. More importantly,  in the intermediate regimes, the force extension typically depends on the nature of the interactions among various domains of the polymer, excluded volume, electrostatic interactions, and ionic concentrations.  These interactions can lead to distinct elastic behaviors.  

Previous investigations have examined the force–extension behavior of neutral polymers, polyelectrolytes, and zwitterionic polymers \cite{radhakrishnan2019force,khan2003monte,panwar2009counterion,stevens2012simulations,stevens2013simulations,wada2005nonlinear}. These models are typically formulated for homogeneously charged chains and therefore do not capture the complex electrostatic environments that can give rise to the rugged energy landscapes often observed in proteins.
Despite extensive work on neutral and homogeneously charged polymers, the role of electrostatic sequence heterogeneity and charge segregation in determining the nonlinear elastic response, hysteresis, and softening behavior of polyampholytes (PAs) remains poorly understood. PA serves as a simplified model system for intrinsically disordered proteins (IDPs) and other charged proteins. Their conformational and phase behaviors depend intricately on parameters such as net charge, charge sequence or block length, chain rigidity, and ionic concentration \cite{das2013conformations,dignon2020biomolecular,long1998electrophoresis,rumyantsev2021sequence,silmore2021dynamics,devarajan2022effect,palariya2024structural}. The influence of charge sequence and the strength of electrostatic interactions on their elastic behavior remains crucial yet not explored. Addressing this gap could provide a deeper understanding of the PA elasticity, PA hydrogels, and the ruggedness of their energy landscape, offering fundamental insights into intrinsically disordered proteins and guiding the design of novel synthetic materials with properties similar to those of tough soft materials \cite{dong2021programmable,lytle2019designing,li2024design}.

In this study, we investigate the elastic behavior of a PA chain under  stretching and force- relaxation using coarse-grained molecular dynamics simulations.  A diblock PA chain's FEC shows a continuous to discontinuous transition, more like a first-order thermodynamic transition,  on increasing the electrostatic strength $(\Gamma_e)$. We demonstrate here that the FEC of the diblock PA chain exhibits a linear regime, followed by three distinct nonlinear regimes.  More importantly, just before the transition, the extension of the chain grows exponentially with the applied force $f$.  We demonstrate a diblock PA chains exhibit markedly different force–extension responses during stretching and relaxation cycles, resulting in pronounced hysteresis in the cycle. The estimated elastic modulus remains constant and increases with the force. Interestingly, near the transition, the elastic modulus decreases with increasing force; this elastic softening is further followed by elastic stiffening in the strongly stretched limit. 
Furthermore, we analyzed the elastic softening related to the structural folding of the PA chain in regimes that elucidate the elastic softening, which becomes stronger with increasing electrostatic strengths and weaker for smaller block PA or multi-block lengths. 

The structural quantities, such as absolute charge displacement and dipole formation, also highlight the mechanisms underlying hysteresis and sharp structural transitions. Our study is further comprehended by the analytical theory of the diblock PA chain, where we have employed the generalized random-phase approximation (GRPA) to estimate the system's free energy, which also demonstrates a sharp structural transition with stretching force\cite{rumyantsev2021sequence}.
Additionally, we have extended our study of the multi-block PA chain and the biologically relevant systems, where we have investigated the elastic behavior of IDPs, such as LAF1 and DDX4. This considered system demonstrates the sharp-structural transition in the presence of the larger charged blocks' elastic softening regimes in such systems.

 The manuscript is organized as follows. Section II describes the simulation model, including the equations of motion and model parameters. Section III presents all results and is divided into four subsections. In Subsection A, we discuss the mechanical response of a diblock polyampholyte (PA), highlighting its elastic behavior, scaling regimes, elastic modulus, force-induced softening, charge displacement, and dipole formation. Subsection B introduces a theoretical framework for understanding the elastic response of diblock PAs. The influence of charge sequence patterning on the mechanical properties of PAs and intrinsically disordered proteins (IDPs) is examined in Subsections C and D. Finally, Section IV provides a summary of our findings and discusses their broader implications.

}




\section{Simulation Model}
\begin{figure}[t]
\includegraphics[width=\linewidth]{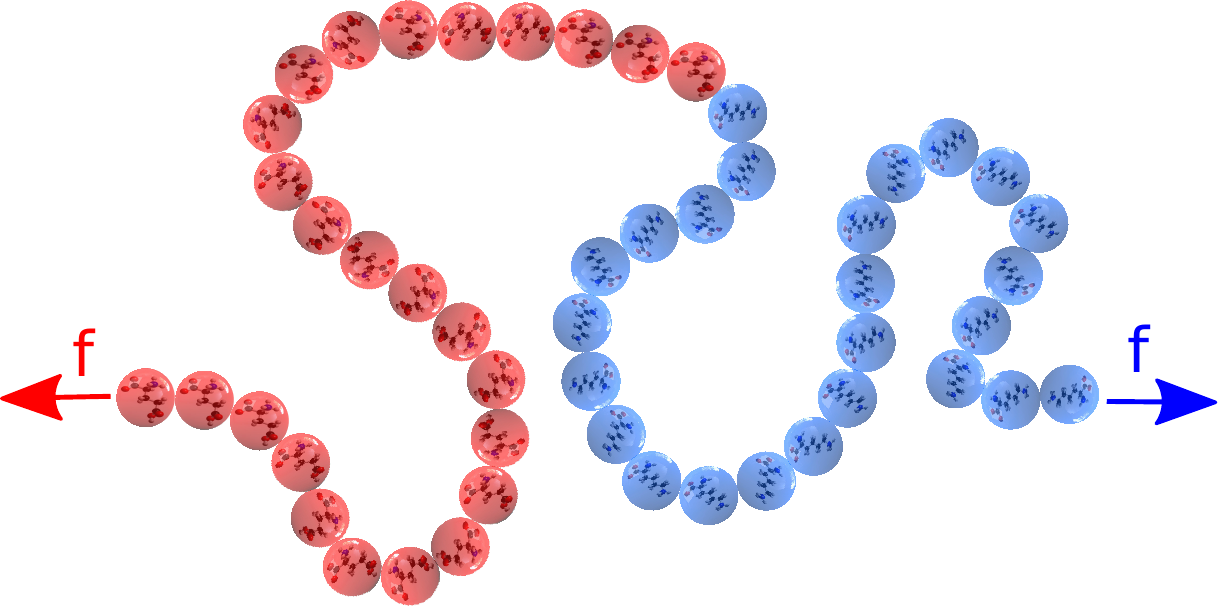}
\caption{A schematic of a diblock PA chain consists of positive(blue) and negative(red) charged monomers, subjected to an external tensile force applied at its terminal monomers in opposite directions. }
\label{Fig:schematic}
\end{figure}

We consider a flexible PA chain composed of an equal number of positive and negative charged monomers such that the net charge of the polymer is zero in a salt-free solution. The monomers of the PA obey the under-damped Langevin equation given as
\begin{equation}\label{Eq:m_i}
    m_i \frac{d^2{\bf r}_i}{dt^2} = - \zeta {\bf v}_i + {\bf F}_i  + {\bf F}_i^t  + {\bm \Gamma}_i(t) , 
\end{equation}
where $m_i$, $\mathbf{r}_i$, and $\mathbf{v}_i$ denote the mass, position, and velocity of the $i^{\text{th}}$ monomer, respectively. The term $-\zeta \mathbf{v}_i$ represents the viscous drag, with $\zeta$ being the drag coefficient. Additionally,  $\mathbf{F}_i$ is the total force acting on the $i^{\text{th}}$ monomer due to interaction potentials, and $\boldsymbol{\Gamma}_i(t)$ denotes the thermal noise, which is Gaussian distributed with zero mean, and its second moment satisfies the fluctuation–dissipation theorem,
$    <{\bm \Gamma}_i(t) \cdot {\bm \Gamma}_j(t')> = 6 \zeta k_BT\delta(t-t')\delta_{ij}$,
where $k_B T$ is the thermal energy. An external tensile force is applied only to the terminal monomers along the $x$-direction,    ${ \bm F}_{i}^t = f(\delta_{iN}-\delta_{i1})\hat x$, which imposes equal and opposite tensile forces on the chain terminals, see Figure~\ref{Fig:schematic}.

The connectivity between adjacent bonds is provided by the harmonic spring potential
\begin{equation} \label{eq:bond_pot}
    U_S = \frac{\kappa_s}{2} \sum_{i=1}^{N-1} \left(|\bm r_{i+1} - \bm r_i| -\ell_0) \right)^2,
\end{equation}
where $\kappa_s$ is the spring constant, $\ell_0$ is the equilibrium bond length, and $N$ is total monomers in the PA. Excluded-volume repulsion is modeled using the truncated and shifted repulsive part of the Lennard–Jones (LJ) potential:
\begin{align} \label{eq:ev_pot}
\displaystyle
U_{LJ} = \left\{ 
\begin{array}{cc} \displaystyle
  4 \epsilon_{LJ}\sum_{i>j}^{N} \left[\left(\frac{\sigma}{r_{ij}}\right)^{12} -\left(\frac{\sigma}{r_{ij}}\right)^{6} + \frac{1}{4}\right], & r_{ij} \leq r_c  \\ \displaystyle
 0 , & r_{ij} > r_c, 
\end{array}
\right. ,
\end{align}
where, $r_{ij}$ is the distance between two monomers, $\epsilon_{LJ}$ is the interaction parameter, and $\sigma$ is the LJ diameter. Here $r_c = \sqrt[6]{2} \sigma$ is the excluded volume cut-off distance between the monomers. Electrostatic interactions among charged monomers are modeled by the long-range Coulomb potential
\begin{equation}\label{Eq:u_e}
U_C= {\frac{1}{4\pi\varepsilon} } 
\sum_{\mathbf{n}}{}^{'} \sum_{i=1}^{N} {\sum_{j=1}^{N}} \frac{z_iz_j e^2}{|\mathbf{ r}_{i,j}+\mathbf{n}L_{pbc}|}. 
\end{equation} 
where $\varepsilon$ is the dielectric constant, which for simplicity is assumed to be homogeneous; $z_i = \pm e$ denotes the charge valency, and $L_{pbc}$ is the length of the primary periodic simulation box. The first summation runs over all periodic images of the primary box, except for the one that avoids self-interactions, i.e., ${\mathbf{n}}(n_x,n_y,n_z) = (0,0,0)$. The long-range Coulomb interaction is computed using the Ewald summation technique \cite{kolafa1992cutoff,toukmaji1996ewald,pollock1996comments,deserno1998mesh}. The total interaction potential of the PA chain is written as $U=U_S +U_{LJ}+U_C$ and the corresponding interaction force is given by ${\bm F}_i=-\nabla_i U$.

{\it Parameters:} All simulation parameters in this manuscript are presented in reduced units. Lengths are expressed in units of the equilibrium bond length $\ell_0$, energies in units of $\epsilon_{LJ}$, and time in units of $\tau = \sqrt{m \ell_0^2 / k_B T}$. The Lennard–Jones (LJ) diameter is set to $\sigma/\ell_0 = 1$, and the thermal energy is normalized such that $\epsilon_{LJ}/k_B T = 1$. The strong harmonic spring constant is chosen $\kappa_s = 10^4 k_B T/\ell_0^2$ to prevent bond stretching.

The Ewald summation method is employed to compute electrostatic interactions, with convergence parameters chosen to ensure that the error in force evaluation remains on the order of $10^{-4}$. Counterion effects on the PA conformations are negligible at dilute concentrations; hence, explicit counterions are omitted. Simulations are primarily performed for chains of length $N = 200$ with block lengths ranging from diblock to alternating sequences. Additional results are reported for selected IDP protein sequences at $N = 236$ and $N=168$, and theoretical comparisons are made for the diblock PA chain at $N = 1000$.

The strength of electrostatic interactions is characterized by the dimensionless parameter $\Gamma_e = \ell_B/\ell_0$, where the Bjerrum length $\ell_B = e^2/(4 \pi \varepsilon k_B T)$ represents the distance at which the electrostatic interaction between two unit charges equals the thermal energy $k_B T$. We consider a range of $\ell_B \in (0.1, 5)$, which spans the experimentally accessible regime, for instance, $\ell_B \approx 0.7 ,\text{nm}$ in water.

The equation of motion is integrated using the velocity–Verlet algorithm with a time step of $\delta t = 10^{-3}\tau$. Each simulation is performed for a total duration of $t = 10^5 \tau$, with the first $5 \times 10^4 \tau$ discarded for equilibration. Statistical averages are obtained from at least 20 independent simulations to ensure reliability.

\section{Results}

\begin{figure}[t]
\includegraphics[width=\columnwidth]{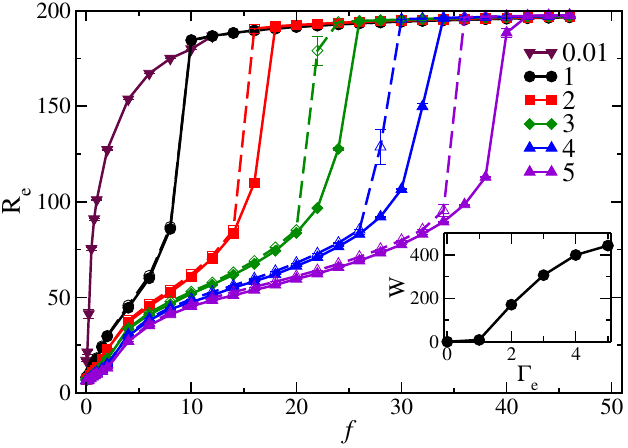}
\caption{Figure illustrates a force-induced, first-order-like discontinuous conformational transition of the diblock PA chain under a constant stretching force. The end-to-end distance is plotted for various electrostatic interaction strengths, $\Gamma_e$, for both the force-extension (solid lines) and force-relaxation (dashed lines). The inset displays the energy dissipation during the force-extension and relaxation cycles. }
\label{Fig:end_distance_lb}
\end{figure}

\subsection{Stretching of diblock PA chain}

A diblock PA chain tends to adopt a compact globular conformation even at relatively low electrostatic strengths, owing to the segregation of oppositely charged monomers into distinct blocks \cite{das2013conformations,sawle2015theoretical,danielsen2019molecular,dinic2021polyampholyte,palariya2024structural}. To characterize the mechanical response of the diblock PA chain under external stretching force, we examine the root-mean-square end-to-end distance, ${R_e} = \sqrt{\langle ({\bm r}_N - {\bm r}_1 )^2 \rangle}$, as a function of the externally applied force $f$ on its terminal ends. Figure~\ref{Fig:end_distance_lb} shows the corresponding force–extension curve (FEC) of the chain for a range of electrostatic strengths ($\Gamma_e$ = 0.01–5).
The solid lines represent the force-extension process, during which the chain is gradually stretched by increasing the applied force at both ends, whereas the dashed lines depict the subsequent force-relaxation process, where the applied force is progressively reduced. During the force-increasing stage, chain conformations obtained at lower forces are used as initial configurations, which are then equilibrated over a sufficiently long simulation time before analyzing $R_e$ and other relevant physical quantities. An identical procedure is employed during the force-relaxation process.


 Our simulations focus on the electrostatic correlation range $\Gamma_e>0.5$,  where the diblock PA chain typically acquires a globular-like structure in the absence of an external force.
 A weak constant force to the terminal ends causes a monotonic swelling of the diblock PA chain.    With increasing force, an abrupt change in the force-extension curve of the PA chain emerges, as shown in Figure~\ref {Fig:end_distance_lb}. This sharp conformational change is characteristic of a phase transition similar to a first-order transition.  At higher stretching forces, an increase in $R_e$ is very slow, as the chain approaches its maximum extension, indicating stiffening under tension. For larger $\Gamma_e$, as expected, the transition shifts to higher critical forces while retaining all the qualitative features of the force–extension response. 
 
 In the limit of small electrostatic strength, $\Gamma_e \ll 1$, the PA chain adopts coil-like conformations. 
The application of a stretching force induces a continuous, smooth, and monotonic extension of the chain. 
At large forces, the chain extension converges to values comparable to those observed at higher $\Gamma_e$.

Next, we examine the force–relaxation behavior of the PA chain. For lower electrostatic strengths ($\Gamma_e \leq 1$), the force–extension and force–relaxation curves coincide, indicating complete reversibility of the process. Interestingly, for higher values ($\Gamma_e > 1$), the sharp structural transition of the chain occurs at a lower critical force compared to the force–extension case. This trend is consistent across all $\Gamma_e$. The deviation between the force–extension and force–relaxation curves is in the vicinity of the transition point, while the remaining curves coincide for both processes for all $\Gamma_e$, as shown in Figure~\ref{Fig:end_distance_lb}.

The deviation between the force–extension and force–relaxation curves arises from strong electrostatic attraction between oppositely charged blocks, which stabilizes a compact globular state. In force relaxation, the PA chain remains extended even at slightly reduced forces, as the external force resists re-association of charged domains.   The inter-block attraction requires a larger force to trigger unfolding, resulting in a hysteresis loop in the extension and relaxation process. The hysteresis domain grows systematically with the electrostatic strength, $\Gamma_e$.

We estimate the energy loss over a complete force–extension and relaxation cycle by evaluating the difference between the areas under the respective curves.  The algebraic expression of the energy loss $W$ is defined as 

\begin{equation}
W=\oint {\bm f}\cdot d{\bm R_e}.    
\end{equation}
The line integral over a closed loop indicates the integration over the extension and relaxation process. 
The corresponding values of $W$ are shown in the inset of Figure~\ref{Fig:end_distance_lb}. At low electrostatic strengths ($\Gamma_e < 1$), the energy loss is nearly negligible, whereas for larger values ($\Gamma_e > 1$), it increases substantially.  This signifies enhanced energy dissipation and highlights the critical role of electrostatic interactions in governing the elastic response of the diblock PA chain. The observed hysteresis is typically attributed to the association and dissociation of physical crosslink bonds during the stretching and relaxation process. Furthermore, it indicates that such a system can sustain higher tension during stretching.

\begin{figure}[t]
\includegraphics[width=\columnwidth]{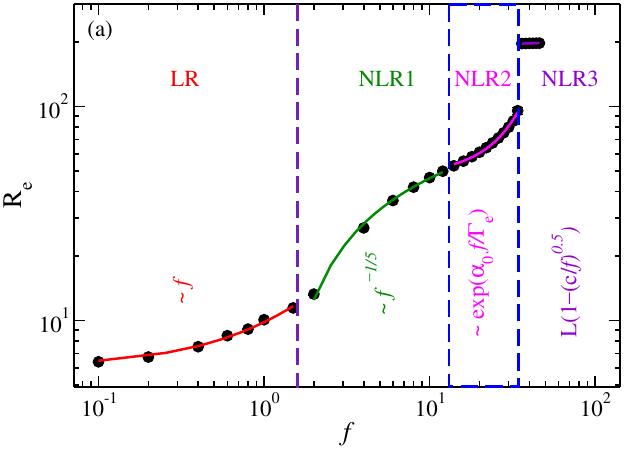}
\includegraphics[width=\columnwidth]{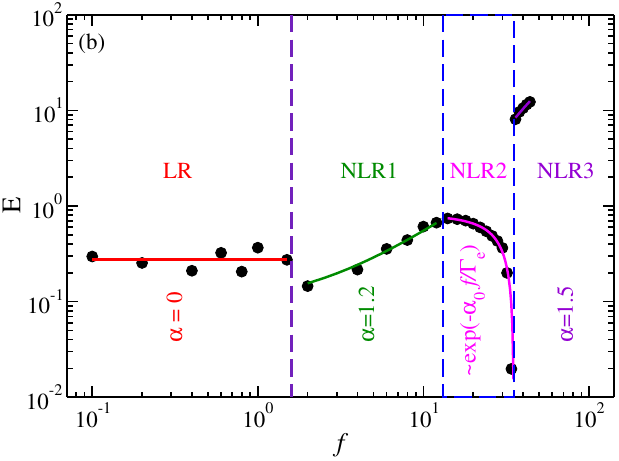}
\caption{Figure (a) illustrates four different elastic regimes of the diblock chain. It shows one linear and three distinct non-linear regimes observed in the force-extension curves, and (b) the corresponding variation in elastic modulus $\mathrm E$ as a function of the external force $f$ at a fixed $\Gamma_e = 5$ and $N=200$.}
\label{Fig:scaling}
\end{figure}

\subsubsection{ Scaling regimes of a diblock PA chain}
 
 The force-extension curve of the PA chain can be categorized into four distinct regimes, based on distinct power-law relations between force and extension.  To elucidate various scaling behaviors, we specifically analyze the force-extension response at $\Gamma_e = 5$ as shown in Figure~\ref{Fig:scaling}a. In the weak force limit $f/\Gamma_e <1$, the extension of the chain grows linearly with applied force at its ends, $R_e  \sim f$, which is referred to as the linear response regime  (LR). This is consistent with the entropic elasticity of an ideal polymer whose elasticity is given by entropic contributions \cite{marko1995stretching,cocco2003force}.

\begin{figure}[t]
\includegraphics[width=0.985\columnwidth]{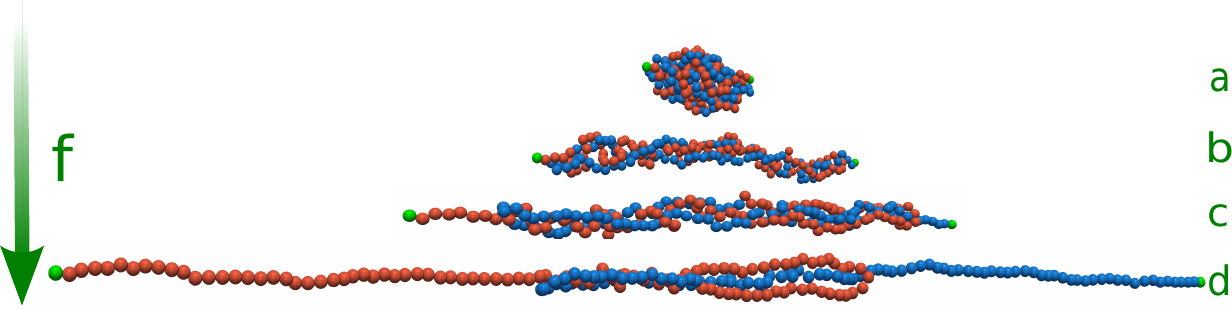}
\caption{ Various conformations of the PA chain under external force. Snapshots demonstrate structural transitions of a diblock PA chain under stretching force for various $f=2,10,20,$ and $38$ at $\Gamma_e=5$ and $N=200$.}
\label{Fig:snap_db}
\end{figure}

The linear regime of the force–extension curve (FEC) is followed by two distinct nonlinear regimes, denoted as nonlinear regime 1 (NLR1) and nonlinear regime 2 (NLR2). These regimes arise from strong electrostatic interactions between oppositely charged blocks. In NLR1, the diblock PA chain remains globular while its terminal segments are gradually extended, transforming the chain from a compact to an elongated yet folded state, see Figure~\ref{Fig:snap_db}a and b. During this process, the FEC exhibits a nonlinear dependence characterized by $R_e \sim f^{-1/5}$.

In the NLR2, the extension increases most rapidly with applied force, indicating that chain stretching becomes energetically more favorable. The PA adopts a distinctive doubly folded hairpin-like morphology, where a folded domain persists near the center, and both ends extend outward from it (see Figure~\ref{Fig:snap_db}c and d). In this regime, the end-to-end distance follows an exponential relation with force, $R_e \sim \exp(\alpha_0 f/\Gamma_e)$.

In the limit of very high force $f/\Gamma_e \gg1$, the chain nearly approaches full extension. In this regime, the extension force overpowers the entropic and electrostatic interactions. Therefore, the response follows more like a worm-like chain (WLC) scaling\cite{marko1995stretching}, irrespective of the strength of electrostatic interactions, as shown in  Figure \ref{Fig:end_distance_lb} and Figure~\ref{Fig:scaling}a.    The force extension in the WLC model  can be expressed 
\begin{equation}
   f\frac{ \ell_K}{\ell_0} \approx \frac{2R_e}{L} + \frac{L^2}{2(L - R_e)^2} - \frac{1}{2}, 
\end{equation}
  where $L = (N-1)\ell_0$  is the chain length and  $\ell_K$ is the Kuhn length. This relationship in the high force limit is approximated 
  \begin{equation}
  R_e \approx L \left(1 - \sqrt{\frac{a}{f}} \right),
  \end{equation}
where $a$ is some constant.  The relationship shows excellent agreement with the simulation results in the strong force limit, as illustrated in Figures ~\ref{Fig:end_distance_lb} and \ref{Fig:scaling}a.



\subsubsection{Elastic Modulus}
From the FEC, the elastic modulus ($E$) of a diblock PA chain can also be evaluated as
\begin{equation}
E = \left( \frac{dR_e}{df} \right)^{-1}.
\end{equation}
Figure~\ref{Fig:scaling}b presents the variation of $E$ with the applied force $f$ for $\Gamma_e = 5$, revealing distinct mechanical regimes consistent with the features observed in the FEC. In the low-force limit, the linear dependence of extension on force indicates that $E$ remains constant, following the scaling ${E} \sim f^{\alpha}$ with $\alpha \approx 0$. At large forces ($f / \Gamma_e \gg 1$), the scaling exponent increases to $\alpha  \approx 1.5$, closely matching the worm-like chain (WLC) behavior and corresponding to the NLR3 regime. These linear (LR) and nonlinear (NLR) regimes are clearly marked in Figure~\ref{Fig:scaling}b.

\begin{figure}[t]
\includegraphics[width=\columnwidth]{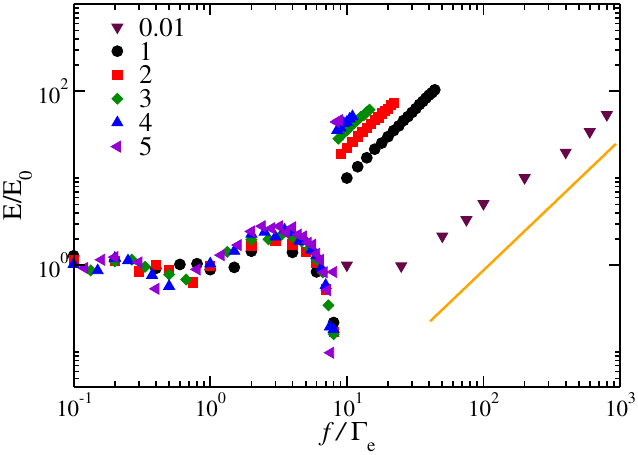}
\caption{The variation of the normalized elastic modulus $E/E_0$ as a function of scaled external force $f/\Gamma_e$ of the diblock PA chain at various electrostatic strengths $\Gamma_e$ from $0.01-5$ for the case of force-extension. The solid lines represent the power-law variations in region NLR3 with exponents $\alpha=3/2$. }  
\label{Fig:stiffness}
\end{figure}

In the intermediate-force regime, the diblock PA chain exhibits pronounced stiffening as electrostatic interactions resist further extension. The end-to-end distance increases nonlinearly with applied force, and correspondingly, the elastic modulus grows following a scaling relation $E \sim f^{\alpha}$ with $\alpha \approx 1.2$, characteristic of the nonlinear stiffening regime (NLR1). Near the critical force, however, the elastic modulus displays an anomaly: $E$ drops sharply and becomes smaller than its equilibrium (zero-force) plateau value. The force-induced softening reflects the strong influence of electrostatic correlations, see Figure~\ref{Fig:scaling}b and Figure~\ref{Fig:stiffness}. Importantly, such softening is absent in the weak-correlation regime ($\Gamma_e \ll 1$).

The origin of this softening of the elastic modulus lies in the dissociation of oppositely charged monomer pairs within the folded hairpin-like conformations, see Figure~\ref{Fig:snap_db}c. As the chain is stretched, these electrostatic bonds dissociate, leading to structural reorganization into partially folded and elongated configurations. In these conformations, oppositely charged blocks in the central region rearrange into a three-layered symmetric pattern of positive and negative domains, followed by stretched single arms on both sides, as shown in Figure~\ref{Fig:snap_db}d. This structure is mechanically fragile, as the central axis contains alternating charged segments that promote further elongation along the x-axis. Consequently, the chain exhibits a significant reduction in its elastic modulus.

Figure~\ref{Fig:stiffness} illustrates the variation of the normalized elastic modulus, $E/E_0$, as a function of applied force for different electrostatic strengths ($\Gamma_e$). In the strong-correlation regime $\Gamma_e > 1$, the diblock PA chain exhibits all four distinct mechanical regimes discussed earlier. Remarkably, when the applied force is rescaled by $\Gamma_e$, the curves corresponding to different electrostatic strengths collapse onto a single master curve over the first three regimes—namely, the linear response, nonlinear regime 1, and elastic softening regime. This collapse of various curves reveals universal scaling behavior, implying that the dimensionless ratio $f/\Gamma_e$ serves as a single governing parameter for the mechanical response within the globular states.
In contrast, in the weak-correlation regime ($\Gamma_e \ll 1$), force-induced softening is absent, highlighting that this behavior is a distinctive feature of strongly correlated PA chains. 


\subsubsection{Elastic Softening}
Figure~\ref{Fig:snap_db} c and d  illustrate the associated conformational states of the diblock PA chain in the elastic softening regime.
To understand the origin of the elastic-softening behavior, we now focus on the electrostatic energy, which is a major contributor to the PA chain's total energy, particularly in this regime.    We make a heuristic approximation of the conformation shown in Figure~\ref{Fig:snap_db}d  as six rod-like segments arranged in parallel, as shown in Figure~S1.  The associated electrostatic energy of this conformation is computed within the limit assuming that the $R_e, L > 2s_0$, where $2s_0$ is the overlap region of the segments.  The approximate form of the electrostatic energy of these conformations is given as a function of $R_e$

\begin{widetext}
   \begin{equation}\label{Eq:total_elec_re}
       U_e(R_e)/\Gamma_e \approx  4 + L/2 - \frac{20}{L - R_e} - 2R_e + \frac{R_e^2}{2L} + R_e \log [8] - L \log [16] + R_e \log \left[\frac{R_e}{2}\right] ,
\end{equation}     
\end{widetext}

Ignoring higher-order terms and keeping only those terms that depend on $R_e$, we can write down the electrostatic energy as  
\begin{equation}\label{Eq:total_elec_re1}
       U_e(R_e)/\Gamma_e \approx   - 2R_e  - \frac{20}{L - R_e}+  \frac{R_e^2}{2L} + R_e \log [8]  + R_e \log \left[\frac{R_e}{2}\right].
\end{equation} 

The derivative of  the electrostatic energy w.r.t. $R_e$ yields the  force which is equivalent to the tensile force $f$, 
\begin{equation}\label{Eq:tensile_force}
    f(R_e)/\Gamma_e \approx \frac{R_e}{L} -  \frac{20}{(L - R_e)^2} +  \log\left[\frac{R_e}{2}\right].
\end{equation}
 The second derivative of $U_e$  with $R_e$ can corresponds to the elastic modulus,
 \begin{equation}\label{Eq:elasticity}
    E(R_e)/\Gamma_e \approx  \frac{{1}}{L}  - \frac{40}{(L - {R_e})^3} + \frac{1}{R_e}.
\end{equation}
Assuming $R_e/L<1$, we can make an approximation here to find the behavior of the function $R_e\sim \exp(\alpha_0 f/\Gamma_e)$, with the $\alpha_0$ some constant. Using this approximate relation, we obtain a qualitative dependence of the elastic modulus as a function of the tensile force $f$, 
\begin{equation}
E\sim  E_0+ \alpha_1\exp(-\alpha_0f/\Gamma_e).     
\end{equation}
The above expression clearly indicates that the elastic modulus exponentially decreases with $f$ for the conformations used. 

In summary, we have demonstrated that the elastic softening regime is a consequence of the structural transition from a globular structure to a double hairpin loop, which exhibits the lower elastic modules.  This analysis provides a clear insight into the connection between the microscopic interactions between charged domains and the macroscopic elastic response of the diblock PA chain under tension.


\begin{figure}[t]
\includegraphics[width=\columnwidth]{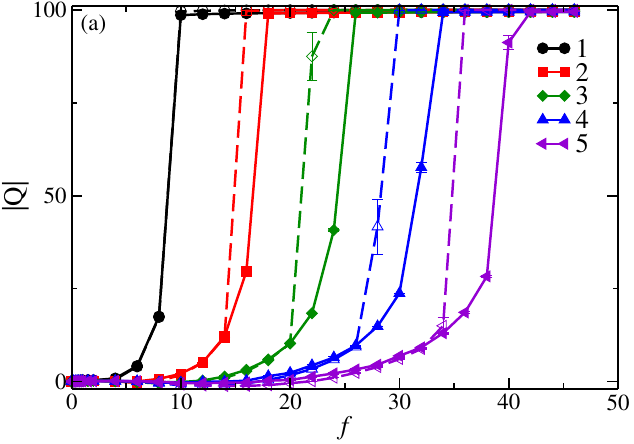}
\includegraphics[width=\columnwidth]{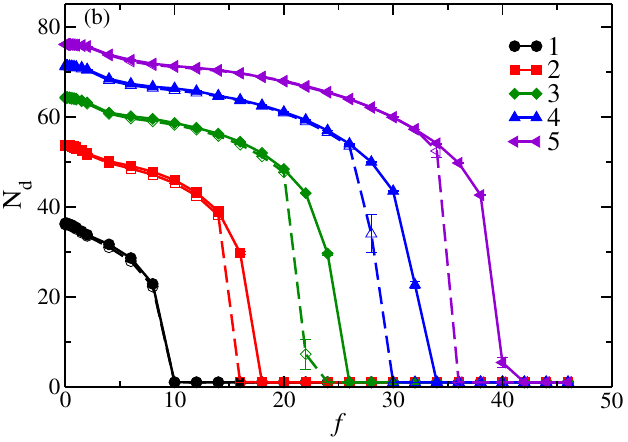}
\caption{The variation of the absolute charge displacement $|Q|$ from the center of mass to one end of the diblock PA chain (a) and the number of dipoles (b) as a function of $f$ at  $\Gamma_e$ from $1-5$. The solid lines correspond to the case of force-extension, and dashed lines are for the force-relaxation.}
\label{Fig:abs_charge}
\end{figure}

\subsubsection{Charge Segregation and Dipole Formation}
To gain further insight into the origin of the hysteresis, we analyze the charge distribution along the direction of the external force by calculating the absolute charge displacement, $|Q|=|<\sum_{i=1}^{N_x} z_i>|$ measured from the chain’s center of mass along the $x$-direction. Here, $z_i = \pm 1$ denotes the valency of the $i^{\text{th}}$ monomer, and $N_x$ is the number of monomers on one side of the center of mass. The quantity $|Q|$ characterizes the degree of charge segregation along the chain and ranges from $0$ for a homogeneous globular state to a maximum value of $N/2$ for a fully stretched configuration ( see Figure~\ref{Fig:abs_charge}a).

For weak electrostatic interactions $\Gamma_e \leq 1$, $|Q|$ increases monotonically from $0$ to $N/2$ with  $f$, and the force–extension and force–relaxation curves overlap, indicating reversible conformational transitions and negligible hysteresis. As the electrostatic strength increases, $|Q|$ grows smoothly with $f$ before displaying an abrupt jump near the critical force, corresponding to a charge segregation transition. Notably, the average charge displacement, $|Q| \approx 12$, after the abrupt jump in conformations is independent of the strength of electrostatic correlations for all $\Gamma_e$.
The asymmetry in $|Q|$ between extension and relaxation provides a direct structural signature of hysteresis: during extension, charges are more effectively segregated away from the chain center, whereas upon relaxation, the system retains a partially mixed charge distribution.


In addition to charge displacement, we quantify the formation of effective dipoles within the diblock PA chain (Figure~\ref{Fig:abs_charge}b). The number of dipoles, $N_d$, is defined as the total count of nearest-neighbor pairs of oppositely charged monomers located within a cutoff distance of $1.25\sigma$, excluding consecutive monomers along the backbone. Each monomer can participate in only one dipole pair; once it is counted as part of a dipole, it is excluded from forming additional oppositely charged pairs. This provides a microscopic indicator of local electrostatic associations, which stabilize compact globular and hairpin-like conformations.

At low electrostatic strengths, the force–extension and force–relaxation curves nearly coincide, indicating negligible hysteresis and minimal energy dissipation. As the applied force increases, $N_d$ decreases monotonically and drops sharply to zero near the critical force, marking the complete dissociation of dipolar pairs. In the strong-correlation regime $\Gamma_e > 1$, a pronounced hysteresis emerges in the $N_d$–$f$ curves, with the number of dipoles during force relaxation remaining significantly higher than during force extension, see Figure~\ref{Fig:abs_charge}(b). The overall magnitude of $N_d$ increases with $\Gamma_e$, reflecting the enhanced probability of oppositely charged segments forming folded or globular states. This greater dipole number effectively induces additional energy barriers against unfolding, thereby amplifying the hysteresis and increasing energy dissipation during cyclic stretching.



\subsection{Theory of Diblock PA Chain Under Extension }
We present a theoretical framework to quantify the elastic response of a diblock PA chain subjected to a constant extensile force at its terminal ends. Electrostatic interactions among monomers are treated within the generalized RPA (GRPA) framework, which accurately captures electrostatic correlations in both the weak  ($\Gamma_e<1$) and strong coupling regimes ($\Gamma_e > 1$) for the neutral PAs. The total free energy of the diblock PA chain is written as
\begin{equation}\label{eq:free_energy}
\mathcal{F}(R_g) = \mathcal{F}_{C}(R_g) + \mathcal{F}_{V}(R_g) + \mathcal{F}_{E}(R_g) + \mathcal{F}_{T}(R_g),
\end{equation}
where $\mathcal{F}_{C}$, $\mathcal{F}_{V}$, $\mathcal{F}_{E}$, and $\mathcal{F}_{T}$ denote contributions from chain connectivity, volume exclusion, electrostatic correlations, and external tensile force, respectively. 

The entropic part of free energy consists of the confinement and stretching  contribution\cite{rubinstein2003polymer,rumyantsev2022unifying},
\begin{equation}\label{eq:free_conf}
    \beta \mathcal{F}_{C}(R_g) = \frac{9}{4}\left(\frac{6 R_g^2}{N \ell_0^2} + \frac{N \ell_0^2}{6 R_g^2} \right) , 
\end{equation}
where $\ell_0$ is the bond length of the diblock PA chain.

The short-range repulsion due to the excluded volume interactions in a globule of volume $V$ is computed assuming polymer monomers are hard spheres of diameter $\sigma$ at packing fraction  $\eta$.  Therefore, we use the Carnahan–Starling equation of state to compute the free energy relation, which is best fitted for hard spheres at larger densities \cite {mcdonald2006theory,zhang2016salting,carnahan1969equation,zaccarelli2009crystallization}.  Thereby, the excluded volume contribution can be expressed as
\begin{equation}\label{eq:free_vol}
   \beta \mathcal{F}_{V}(R_g) = \frac{V}{\sigma^3} \phi \left[  \frac{4 \eta - 3\eta^2}{(1 - \eta)^2} - 4\eta \right]  ,
\end{equation}

where $\eta = {\pi \phi}/{6}$, $\phi = {N\sigma^3}/{V}$, and $V = 4 \pi R_g^3/3$. The last term in the free energy ensures compatibility with $\Theta$-solvent conditions\cite{rumyantsev2022unifying}.


The contribution of electrostatic interactions is expressed in the GRPA framework, which incorporates the discreteness of charge along the PA chain and a finite number of fluctuating modes corresponding to charge-density fluctuations. The $\beta \mathcal{F}_{E} $ is expressed as
\begin{align}\label{eq:free_coul}
    \beta \mathcal{F}_{E}(R_g) &= \frac{V}{\sigma^3} F_{\mathrm{RPA}}, \\[6pt] 
    F_{\mathrm{RPA}} &= \frac{\sigma^3}{2} \int_{0}^{q_0} \frac{d\mathbf{q}}{(2\pi)^3} 
    \left[ \ln\!\left(1 + \frac{g_{\mathrm{db}}(q)}{(r_D q)^2} \right) 
    - \frac{g_{\mathrm{db}}(q)}{(r_D q)^2} \right],\nonumber
\end{align}
where $r_D = (4 \pi \Gamma_e z_f \phi)^{-1/2} \sigma$ is the Debye screening length for the system of free charges and  $z_f$ is the fraction of charged monomers.   The complete details of the derivation of the free energy can be found in Ref.~\cite{rumyantsev2022unifying}. 
The electrostatic correlation-free energy is calculated using the random phase approximation for the diblock PA, which requires the single-chain structure factor $g_{\mathrm{db}}(q)$ of the PA chain as input in the theory.
For diblock PA chain  the structure factor is approximated as ${\mathrm g}_{db}(q) \approx 1 + {12 z_f}/{(q\sigma)^2}$.   Rumyantsev {\it et. al. }\cite{rumyantsev2022unifying} proposed in the  GRPA  to exclude the non-physical modes of the charge density fluctuation by considering the cut-off wavevector 
$q_0 = (9 \pi^2 z_f \phi)^{1/3} \sigma^{-1}$\cite{rumyantsev2022unifying}.  Integrating Eq.~\ref{eq:free_coul}, the RPA free energy is obtained, which is also valid in the strong correlation limit,

\begin{widetext}
  \begin{equation}\label{eq:rpa}
    F_{RPA} = \frac{1}{12 \pi^2 \left(\frac{r_D}{\sigma}\right)^3}
\Bigg[t_0^3 \ln\left(1 + \frac{1}{t_0^2} + \frac{1}{s^2 t_0^4}\right)- t_0 - \frac{(1-\chi)^{3/2}}{\sqrt{2}} 
\arctan\!\left(\frac{\sqrt{2}\, t_0}{\sqrt{1-\chi}}\right)\\
 - \frac{(1+\chi)^{3/2}}{\sqrt{2}} 
\arctan\!\left(\frac{\sqrt{2}\, t_0}{\sqrt{1+\chi}}\right)
\Bigg]
\end{equation}  
\end{widetext}

$t_0 = r_D q_0$, $s = \sqrt{\pi \phi \Gamma_e/3}$, and $\chi^2 = 1 -4s^{-2}$.  Note that in the above expression (Eq.~\ref{eq:rpa}), a term $3/(s^2 t_0)$ is subtracted, which arises from the chain connectivity.  

The last term of the total free energy is due to the tensile force applied at the ends of the PA chain 
\begin{equation}\label{eq:free_tensile}
    \beta\mathcal{F}_{T}(R_{\mathrm g}) = -fR_e, 
\end{equation}
Here, we take $R_e \approx 2 R_g$ to close the free energy expression in terms of the radius of gyration. This closure relation is consistent with our simulation results and provides a mapping between the two length scales.


The equilibrium radius of gyration ($R_g$) is obtained numerically by minimizing Eq.~\ref{eq:free_energy} with respect to $R_g$ at a given stretching force $f$. The resulting equilibrium values of $R_g$ are shown in Figure~\ref{Fig:theory_rg}a as a function of $f$ for different $\Gamma_e$. Interestingly, the GRPA theory predicts a continuous transition in the weak-correlation regime, whereas for $\Gamma_e \geq 0.05$, it demonstrates a discontinuous change beyond a critical stretching force, characteristic of a globule–coil transition in the strong-correlation regime. This transition is identified as a first-order conformational phase transition of the PA chain. Notably, as expected,  the transition point shifts for the larger critical force $f_c$ for higher electrostatic strengths $\Gamma_e$. 


The presented theoretical predictions are in good qualitative agreement with the trends observed in our simulations, confirming the model's validity in capturing both continuous and first-order discontinuous transitions.  




\begin{figure}[t]
\includegraphics[width=\columnwidth]{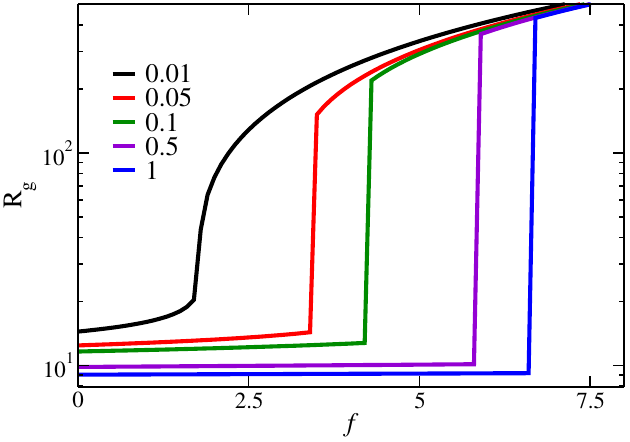}
\caption{  The figure illustrates a continuous transition of the PA chain under a constant stretching force for weak electrostatic correlations ($\Gamma_e = 0.01$), and a discontinuous transition in the globular phase for stronger correlations ($\Gamma_e \geq 0.05$) occurring beyond a critical force $f > f_c$. The radius of gyration $R_g$ of a diblock PA chain as a function of external force $f$ for different $\Gamma_e$ at $N=10^3$. }
\label{Fig:theory_rg}
\end{figure}


\begin{figure}[t]
\includegraphics[width=\columnwidth]{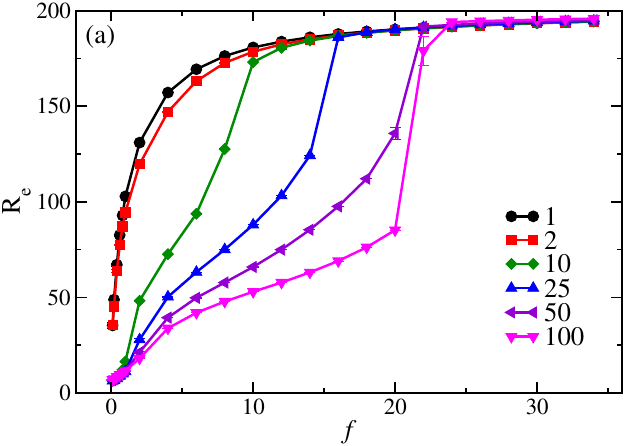}
\includegraphics[width=\columnwidth]{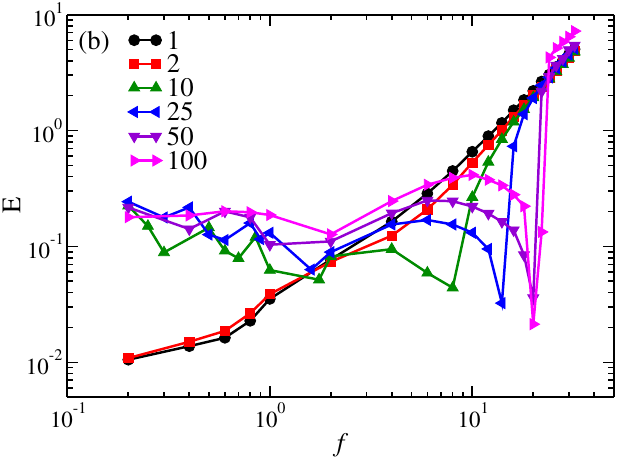}
\caption{ (a) The force-extension curve of PA chain for different block lengths from alternating to diblock chain at $\Gamma_e=3$ and $N=200$, showing the continuous transition for lower block PAs, whereas larger block length PA chain undergoes a discontinuous transition. (b) The variation of the elastic modulus; elastic softening emerges  for the larger block length PAs.}
\label{Fig:end_distance_block}
\end{figure}

\begin{figure}[t]
\includegraphics[width=0.985\columnwidth]{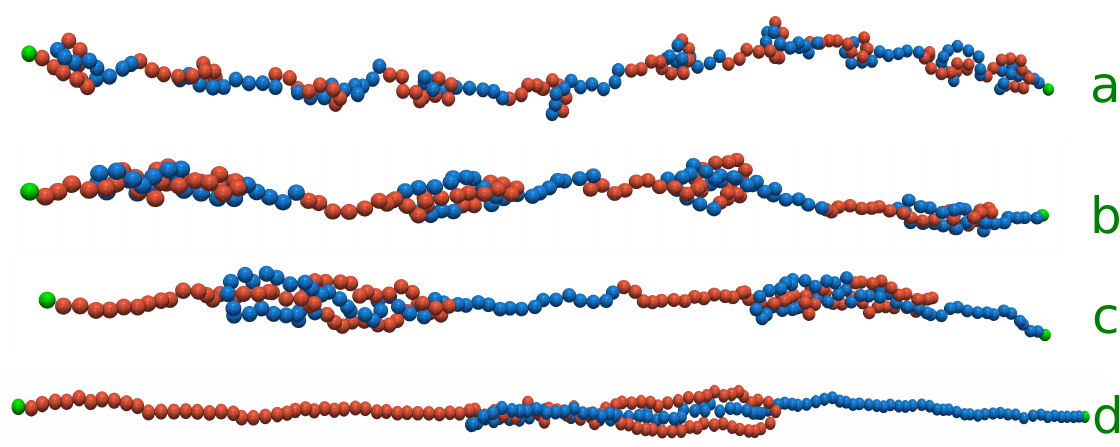}
\caption{ The variation in conformational states of the PA chain for various block lengths with the external force $f$ at   $L_b = 10$ (a), $L_b = 25$ (b), $L_b = 50$(c), and $L_b = 100$ (d). }
\label{Fig:block_snap}
\end{figure}

\subsection{Effect of charge sequence}

In the previous section, we discussed the force-extension behavior of the diblock PA chain, now  we extend this investigation to consider how the charge sequence  of monomers—namely, the block length affects the behavior of FEC. 
Figure~\ref{Fig:end_distance_block}a illustrates the FEC at $\Gamma_e = 3$ for various PAs with varying block lengths, ranging from alternating sequences ($L_b=1$) to the diblock PA chain ($L_b=100$).  For smaller block PAs, the PA chain adopts a coil-like conformation in equilibrium, as the local electrostatic attractions between oppositely charged monomers are effectively screened. Upon application of a stretching force at the chain ends, the chain undergoes a smooth and continuous transition from the coiled state to a fully extended configuration, resembling the classical force-extension response observed in uncharged flexible polymers.

In contrast, increasing block length leads to a decrease in end-to-end distance at weaker force, arising from the overcoming of electrostatic attraction between the opposite monomers over the electrostatic repulsion between the same charged monomers. On the application of an external force to the end monomers, the force–extension curve deviates significantly from that of the small block length PA  chains. Instead of a continuous streching, the PA chain exhibits structural rearrangements into double hairpin-like conformations, requiring a larger force for the fully stretched state, and the number of such double hairpin-like domains increase  while their size decreases. In the case of the diblock PA chain the strong Coulombic attraction between oppositely charged domains produces a compact globular state. As the applied force increases beyond a critical threshold, the chain experiences an abrupt increase in the end-to-end distance. This sharp transition distinguishes diblock PAs from other block PAs, where the unfolding proceeds more gradually.

 Figure~\ref{Fig:end_distance_block}b shows the dependence of the elastic modulus on the charge sequence of the PAs is presented. For lower block length PA chains, the elastic modulus increases monotonically with the applied external force, indicating force-induced stiffening, similar to that of flexible neutral polymers, where entropic elasticity governs the response. In contrast, with the increase in block length, the elastic modulus exhibits a non-monotonic behavior, which arises due to the formation of folded double hairpin-like conformations. The elastic modulus exhibits both stress stiffening and stress softening regimes. The stress-softening increases with the increase in block length as larger block lengths promote the formation of larger-sized folded conformations, and consequently, a larger force is required to attain full contour extension.

\subsection{ Intrinsic Disordered Proteins} To further elucidate the role of charge sequencing, we examine two intrinsically disordered proteins (IDPs), LAF1 and DDX4. LAF1 is localized in P-granules and plays a crucial role in liquid–liquid phase separation and phytochrome-mediated signaling pathways in plant cells, whereas DDX4 is implicated in antiviral defense mechanisms and translation regulation. We employ coarse-grained representations of these proteins comprising $N = 168$ and $N = 236$ residues for LAF1 and DDX4, respectively. Detailed descriptions of the simulation methodology and sequence information are provided in the Supplementary Material \cite{sundaravadivelu2024sequence}.

Figure~\ref{Fig:idp} presents the variation of the elastic modulus for the two IDPs—DDX4 (solid line) and LAF1 (dashed line)—as a function of the charge decoration parameter (nSCD). Smaller nSCD values correspond to mixed charge sequences, whereas larger nSCD values denote well-segregated charge distributions. Sequences with lower nSCD values display a monotonic increase in elastic modulus with increasing force. In contrast, for higher nSCD values, the elastic modulus exhibits force-induced softening, occurring predominantly in variants with larger nSCD. This behavior closely parallels the stress-softening transition observed in polyampholytes with longer block lengths (see Figure~\ref{Fig:end_distance_block}), highlighting the close relation between charge sequence patterning and the mechanical response of disordered polymers.


\section{Conclusions}
We have systematically investigated the elastic response of PA chains with varying charge sequences using coarse-grained molecular dynamics simulations, as well as within the theoretical framework of the GRPA.  At low electrostatic strengths and short block lengths, PA chains undergo a continuous transition from a coiled to a stretched state. In contrast, at higher electrostatic strengths, they form compact globular structures that exhibit an abrupt, first-order-like transition under an external stretching force. The theoretical framework reproduces these transitions across variations in $f$ and $\Gamma_e$, underscoring the robustness of our findings. We further extended this framework to IDPs such as LAF1 and DDX4, whose mechanical responses exhibit similar trends: sequences with lower normalized sequence charge decoration (nSCD) parameters and smaller block sizes show continuous transitions, whereas those with larger nSCD values (longer charge blocks) display abrupt transitions. These behaviors closely mirror those observed in PA chains, suggesting that the underlying mechanisms are universal to sequence-charge-driven systems.

Our study reveals hysteresis in the force–extension and relaxation process indicating enhanced energy dissipation —a hallmark of viscoelastic materials such as hydrogels and biopolymer networks~\cite{hughes2016physics,cluzel1996dna,smith1996overstretching,savin2013two,king2013revealing,pupo2013dna,schwaiger2002myosin}. In such systems, hysteresis typically arises from association and dissociation of physical bonds and the redistribution of mechanical load, contributing to energy loss during cyclic deformation. The force-induced softening has been observed in biomolecular systems such as dsDNA and proteins, where overstretching destabilizes secondary structural organization, especially hydrogen bonding, thereby inducing significant energy dissipation and mechanical softening\cite{hughes2016physics,cluzel1996dna,smith1996overstretching,savin2013two,king2013revealing,pupo2013dna,schwaiger2002myosin}.  

Analysis of the force–extension curves (FECs) reveals four distinct elastic regimes: an initial linear response followed by three nonlinear regions characterized by the elastic modulus. In the low-force regime, the modulus remains constant $E \sim f^0$. At intermediate forces, the chain exhibits stress stiffening $E \sim f^{1.2}$, which is subsequently followed by a stress-softening regime $E \sim \exp(-\alpha_0f/\Gamma_e)$, arising from strong electrostatic coupling. This stress-softening regime is absent at low electrostatic strengths but becomes prominent as block length increases. At higher forces, the chain stiffens again with $E \sim f^{3/2}$~\cite{marko1995stretching,dobrynin2010chains}. The stress-softening behavior is predominantly associated with block PA chains, whereas such effects are generally not observed in neutral polymers or uniformly charged polyelectrolytes. Our analysis reveals that the microscopic origin of the exponential softening is due to electrostatic coupling between oppositely charged folded domains of the double hairpin-like structures. Using structural inputs from this regime, we analytically show that the elastic modulus decays exponentially with applied force. The theoretical framework aligns closely with the simulation behaviors, capturing both continuous transitions in the weakly correlated regime and first-order-like discontinuous transitions under strong electrostatic correlations, thereby validating the simulation findings. 

\begin{figure}[t]
\includegraphics[width=\columnwidth]{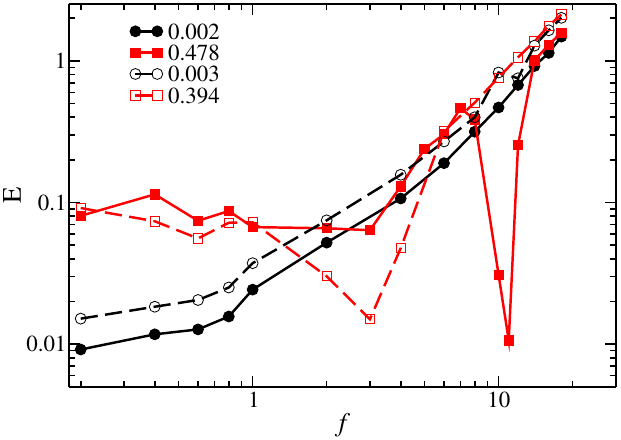}
\caption{The variation of the elastic modulus for different IDPs, specifically DDX4 (solid lines) and LAF1 (dashed lines), at $\Gamma_e = 3$, showing elastic softening in IDPs with higher normalized sequence charge decoration (nSCD). }
\label{Fig:idp}
\end{figure}

Our simulations reveal that the stress-softening regime is observed only in longer block PAs, which adopt globular conformations in the absence of a stretching force, whereas it is absent in shorter block PAs that remain coiled. Together, these results elucidate microscopic origins of nonlinear elasticity and dissipation in PAs and IDPs, establishing a unified framework for predicting their mechanical response across synthetic and biological systems.

In summary, our study reveals new aspects of polymer mechanics resulting from charge segregation and electrostatic correlations.  The mechanical response of a single polymer has paramount relevance in understanding the behavior of polymeric materials, brushes,  networks\cite{rubinstein2003polymer, stabley2012visualizing},  protein functionality, RNA energy landscapes\cite{camunas2016elastic}, and cellular elasticity\cite{storm2005nonlinear,saleh2009nonlinear}.  This provides a direct link between macroscopic elastic behavior and the microscopic structural arrangements of the chain\cite{saleh2009nonlinear}.  More broadly, our findings provide implications, ranging from the mechanical characterization of biological tissues to the rational design of responsive self-healing hydrogels\cite{chen2015novel,long2018salt,li2024design},  tunable soft materials of low-dissipation\cite{zhu2024tough,li2025mechanical}, etc.

\section*{Acknowledgments}
RP and SPS acknowledge financial support from the UGC-India and DST-SERB Grant No. CRG/2020/000661,  and computational time at IISER Bhopal and the PARAM Smriti NSM facility.



\clearpage
\renewcommand{\thesection}{SI-\Alph{section}} 
\setcounter{section}{0}

\addcontentsline{toc}{section}{Supplementary Information}

\twocolumngrid 
\onecolumngrid  

\begin{center}
    \Large\textbf{Supplementary Information}
\end{center}
\vspace{2em}
\twocolumngrid  

\section{Elastic Softening}
\begin{figure}[t]
\includegraphics[width=\columnwidth]{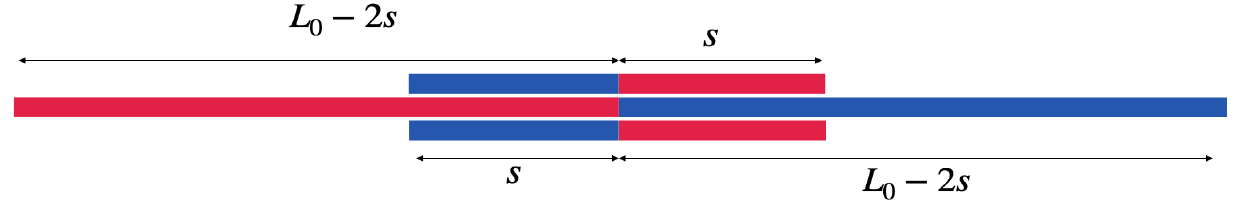}
\caption{The approximate conformation of the PA chain, corresponding to the elastic softening regime prior to the abrupt conformational transition, is illustrated in Figures 5(c) and 5(d) of the main manuscript. The blue corresponds to the positive charge and the red to the negative charge.  }
\label{Fig:sch_th}
\end{figure}

In this section, we approximate the conformation of the PA chains shown in Figures 5(c) and 5(d) of the main manuscript. The schematic representation of this approximate conformation is illustrated in Figure\ref{Fig:sch_th}. We focus on the electrostatic energy of the PA chain, assuming it to be the dominant contribution to the system’s total energy. The PA chain begins to show elastic softening when it adopts conformations similar to those depicted in Figures 5(c) and 5(d) (main manuscript). To evaluate the electrostatic contribution to the total energy, we approximate these conformations  rod-like segments as shown in Figure \ref{Fig:sch_th}. The electrostatic energy can then be computed analytically,
\begin{widetext}
    \begin{equation}\label{Eq:elec}
\begin{aligned}
    U_e = &4 - 2L_0 - 4\sqrt{1 + (L_0 - 3s)^2} + 8\sqrt{1 + (L_0 - 2s)^2} - 4\sqrt{1 + (L_0 - s)^2} + 12\sqrt{1 + s^2} \\
    &\quad - 8\sqrt{4 + s^2} - (L_0 - 2s)\log 4 - 4s \log 4 + 4(L_0 - 3s) \log \left[L_0 + \sqrt{1 + (L_0 - 3s)^2} - 3s\right] \\
    &\quad + 2(L_0 - 2s)\log[L_0 - 2s] - 8(L_0 - 2s)\log\left[L_0 + \sqrt{1 + (L_0 - 2s)^2} - 2s\right] \\
    &\quad + 4(L_0 - s) \log \left[L_0 + \sqrt{1 + (L_0 - s)^2} - s\right] + 4s \log s \\
    &\quad - 12s \log \left[s + \sqrt{1 + s^2}\right] + 4s \log \left[s + \sqrt{4 + s^2}\right] \\
    &\quad + 2s \log \left(\frac{1}{4}\left(s + \sqrt{4 + s^2}\right)^2\right).
\end{aligned}
\end{equation}
\end{widetext}
In this analysis, we assume that the rod-like segments are slender and sufficiently long. By performing a series expansion, the electrostatic energy is simplified and expressed in the following form,
\begin{widetext}
   \begin{equation}\label{Eq:simplified}
   U_e = 4 - \frac{5}{s} + 4s + \frac{4s^2}{L_0} - L_0 \left(2 + \log[4]\right) - s \log[4096] + 2 \left(L_0 - 2s\right) \log[L_0 - 2s].
\end{equation} 
\end{widetext}
The chain length of $L=2L_0$, with the relationship between $s$ and end-to-end distance $R_e$, which can be approximated as $s = (L -R_e)/4$. Therefore, the total electrostatic energy of this configuration can be approximated as
\begin{widetext}
   \begin{equation}\label{Eq:total_elec_re}
       U_e(R_e)/\Gamma_e \approx  4 + L/2 - \frac{20}{L - R_e} - 2R_e + \frac{R_e^2}{2L} + R_e \log [8] - L \log [16] + R_e \log \left[\frac{R_e}{2}\right] ,
\end{equation} 
\end{widetext}
Ignoring higher-order terms and keeping only those terms that depend on $R_e$, we can write down the approximate form electrostatic energy as  
\begin{equation}\label{Eq:total_elec_re}
       U_e(R_e)/\Gamma_e \approx   - 2R_e - \frac{20}{L - R_e}  + \frac{R_e^2}{2L} + R_e \log [8]  + R_e \log \left[\frac{R_e}{2}\right].
\end{equation} 
The derivative of  the electrostatic energy w.r.t. $R_e$ yields the  force which is equivalent to the extensile force $f$, 
\begin{equation}\label{Eq:tensile_force}
    f(R_e)/\Gamma_e \approx \frac{R_e}{L} - \frac{20}{(L - R_e)^2}  + \log\left[\frac{R_e}{2}\right].
\end{equation}
 The second derivative of $U_e$  with $R_e$ corresponds to the elastic modulus,
 \begin{equation}\label{Eq:elasticity}
    E(R_e)/\Gamma_e \approx  \frac{{1}}{L}  - \frac{40}{(L - {R_e})^3} + \frac{1}{R_e}.
\end{equation}
Assuming $R_e/L<1$, we can make an approximation here to find the behavior of the function $R_e\sim \exp(\alpha_0 f/\Gamma_e)$, with the $\alpha_0$ some constant. Using this approximate relation, we obtain a qualitative dependence of the elastic modulus as a function of the tensile force $f$, 
\begin{equation}
E\sim  E_0+ \alpha_1\exp(-\alpha_0f/\Gamma_e).     
\end{equation}
The above expression indicates that the elastic modulus exponentially decreases with $f$ for the conformations used.

\section{Local Stretching and Tension}
To elucidate the microscopic origin of the distinct mechanical regimes observed in the force–extension and elastic modulus profiles, we analyze the local stretching of the chain. For this purpose, the PA chain is partitioned into $N_s$ segments of equal arc-length. The average segmental distance is defined as
\begin{equation}\label{Eq:segment}
    R_n(s)=\sqrt{<({\bm r}_{s+n}-{\bm r}_s)^2>},
\end{equation}
and the corresponding average bond energy of the $n$ segment is computed as,
\begin{equation}\label{Eq:average_bond}
U_S^n= \left<\sum_{s}^{s+n-1}\left(|\bm r_{s+1} - \bm r_s| -\ell_0) \right)^2\right>,
\end{equation}

\begin{figure}[t]
\includegraphics[width=\columnwidth]{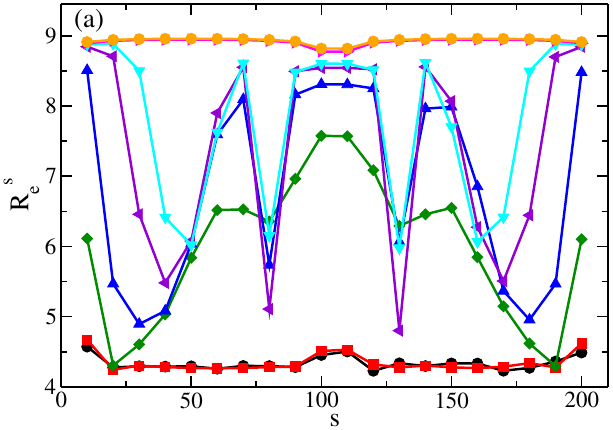}
\includegraphics[width=\columnwidth]{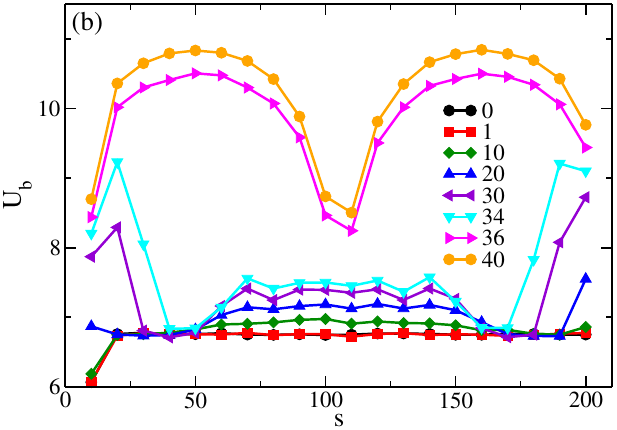}
\caption{The variation of average segmental length(a) and the average bond energy of segments (b) for different force $f$ at fixed $\Gamma_e = 5$ (color coding is the same for both plots) .}
\label{Fig:segment}
\end{figure}

Figure~\ref{Fig:segment} shows the variation of the average segmental length and the average bond energy as a function of stretching force for an arc length $n=10$ at $\Gamma_e=5$, corresponding to the strong correlation regime. At low forces $f=0$ and $f=1$, the variation of the local segment length $R_n(s)$ along the chain is nearly uniform, as illustrated in Figure~\ref{Fig:segment}. This indicates that the applied tension is distributed homogeneously along the chain backbone.  This uniformity is also reflected in the linear relationship observed in Figure. 3 of the FEC (main manuscript). 



As the applied tension increases, segmental extension becomes inhomogeneous; the outer segments of the chain are relatively more elongated, which decreases further along the contour toward the interior. Near the chain center, segments show greater elongation than those at the ends, as seen in Figure~\ref{Fig:segment}a for $ f=10$. With further increase in force, attractive interactions between oppositely charged monomers—particularly pronounced in the inner segments—promote the formation of S-shaped conformations of the positive and negative blocks. Consequently, the segmental length profiles develop a characteristic inverted S-shape pattern, reflecting the alternation of stretched and locally folded domains. These conformations underlie the force-induced elastic softening of the PA chain. Finally, at a high force limit, the PA chain achieves an extended conformation, where segmental lengths once again become nearly uniform across the chain. This regime leads to a stress-stiffening NLS2 regime, and the FEC follows the WLC model.

Figure~\ref{Fig:segment}-b shows the variation of the stretching energy, computed using Eq.~\ref{Eq:average_bond}, within each segment of the PA chain under different applied forces. In the low–force regime, the bond energy is nearly uniform across all segments, consistent with the homogeneous segmental extension discussed earlier. With increasing force, however, the bond energy becomes inhomogeneous: the outer segments have larger elastic energies, while in the inner segments $U_S^n$ decreases for $f=10,20,30$, and $34$, i.e., below the critical force $f_c$. This reduction indicates that these regions experience an effectively smaller stretching force, leading to a corresponding decrease in $R_S^n$. In contrast, at the central segments, the elastic energy increases again, suggesting the emergence of relatively more stretched domains.

For larger forces $f > 40$, the elastic energy profile along the chain develops a double–parabolic shape. Specifically, the energy grows from the ends toward the center, reaches a maximum, and then decreases again near the interface of oppositely charged monomers. The reduction at both ends and the center can be attributed to electrostatic effects: attractive interactions between oppositely charged monomers lower the central tension in fully stretched conformations, while asymmetric electrostatic repulsions near the chain ends reduce the local elastic energy and thus the tension.

Collectively, the observed variations in segmental bond energies across the different force regimes validate and refine the interpretation derived from segmental length analysis, illustrating how local energetic heterogeneity determines the elastic response of diblock PAs.

\begin{figure}[t]
\includegraphics[width=\columnwidth]{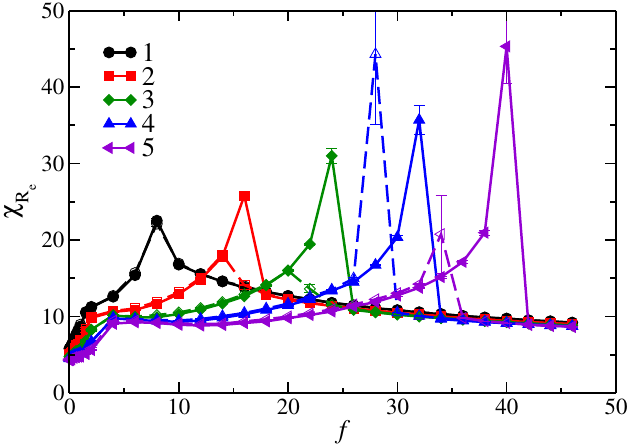}
\caption{Fluctuations in end-to-end distance of a diblock PA chain at various $\Gamma_e$ for the case of force-extension and relaxation behavior at $N=200$.}
\label{Fig:var_re}
\end{figure}

\section{Conformational Fluctuations}
The structural fluctuations associated with the transition points are quantified through the variance  of the end-to-end distance $R_e$, defined as 
$\chi_{R_e} = \sqrt{\langle R_e^2 \rangle - \langle R_e \rangle^2}$, 
as shown in Figure~\ref{Fig:var_re}.  The fluctuations in the structure are relatively small in both the low- and high-force regimes. This gradual increase in fluctuations reflects the continuous nature of the conformational change,  where the chain transitions through intermediate configurations. However, near the critical force, the fluctuations increase significantly, leading to pronounced peaks in $\chi_{R_e}$, which characterize the structural transition. 

In Figure~\ref{Fig:var_re}, the force–extension curves are represented by solid lines, while the dashed lines correspond to force–relaxation. At a weak electrostatic interaction strength  $\Gamma_e = 1$, the stretching and relaxation curves overlap, indicating negligible energy loss and a reversible process. As the electrostatic strength increases, a greater force is required to stretch the PA chain, and the separation between the force–extension and force–relaxation curves becomes more pronounced, reflecting enhanced energy dissipation during the process.

\section{Elastic Modulus from the analytical theory}
We compute the elastic modulus of the polymer in the same parameter regime using the free energy of the PA chain from GRPA given in Eq.18 (main manuscript).  Figure~\ref{Fig:theory_rg} illustrates the variation of the elastic modulus, $E = \partial f / \partial R_{\mathrm g}$, obtained from the GRPA approach. The modulus remains approximately constant over a wide range of forces; however, near the transition point, it exhibits a pronounced drop, followed by a recovery to a slightly lower value compared to the previous saturation level. This signals the mechanical softening of the PA chain under an external stretching force. Beyond the critical force $f_c$, the elastic modulus asymptotically approaches a constant value smaller than the equilibrium limit, in contrast to simulations where additional extension leads to strain hardening and diverges near the maximum extension. These results demonstrate that the theoretical framework quantitatively reproduces not only the globule–coil transition but also the nonlinear elastic response of the chain, specifically the elastic softening of the modulus under constant force.

\begin{figure}[t]
\includegraphics[width=\columnwidth]{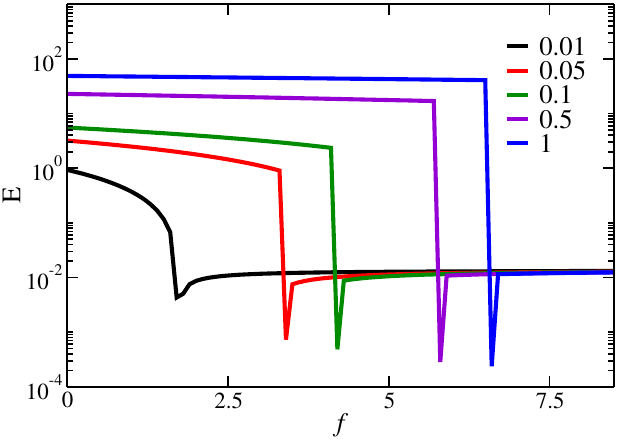}
\caption{  Elastic modulus $E$ of a diblock PA chain, obtained from the GRPA approach using Eq.~18 (main manuscript), shown as a function of the external force $f$ for different values of $\Gamma_e$ at $N = 10^3$.}
\label{Fig:theory_rg}
\end{figure}

\begin{figure}[t]
\includegraphics[width=\linewidth]{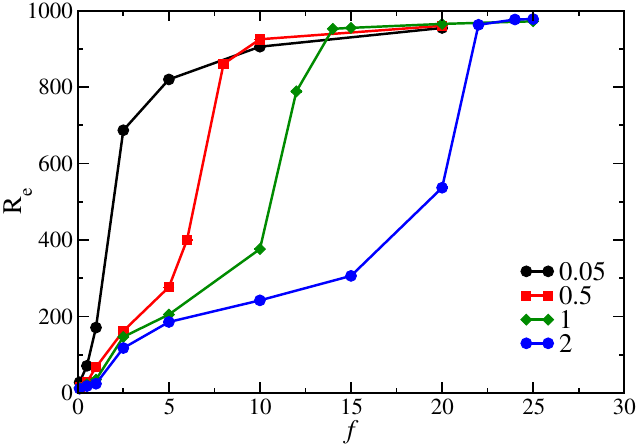}
\caption{  The FEC of a diblock PA chain as a function of external force $f$ for different $\Gamma_e$ and $N=10^3$ from the simulation. The continuous transition for weak correlations for $\Gamma_e=0.05$ at $N=10^3$, and a discontinuous transition in the globule phase occurs for $\Gamma_e>0.05$ beyond a critical force $f>f_c$.  }
\label{Fig:n_1000}
\end{figure}

The presented theoretical predictions Figure.8 (main manuscript), are in good qualitative agreement with trends observed in our simulations\ref{Fig:n_1000} for longe chain lengths to, confirming the validity of the model for capturing both continuous and first-order discontinuous transitions.

\section{IDPs}

To understand the influence of charge patterning, we further investigate naturally occurring IDPs namely  LAF1 and DDX4, consisting of $N=168$, and $N=236$ residues, respectively\cite{sundaravadivelu2024sequence}. For these naturally occurring protein variants, the fraction of charged residues (FCR) and the net charge per residue (NCPR) are $0.262$ and $0.024$ for LAF1, and $0.288$ and $-0.017$ for DDX4, respectively. The charge pattern of each sequence is quantified using the normalized sequence charge decoration (nSCD), which compares each protein’s charge distribution to the maximum and minimum SCD values possible for its composition\cite{rana2021phase,devarajan2022effect}.
\begin{equation}
    SCD = \frac{1}{N} \sum_{i=2}^{N} \sum_{j=1}^{i-1} q_i q_j \sqrt{i-j},
    \label{Eq:SCD}
\end{equation}
where $q_i$ and $q_j$ are charges of the $i^{th}$ and $j^{th}$ monomers, respectively.
The nSCD parameter ranges from 0 to 1, where lower values correspond to well-mixed (alternating) charge sequences and higher values indicate charge-segregated (diblock) sequence\cite{rana2021phase,devarajan2022effect}.

\begin{figure}[t]
\includegraphics[width=\linewidth]{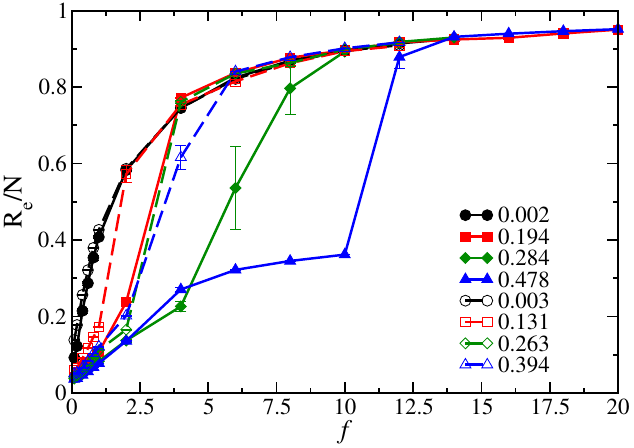}
\caption{The variation of the force extension response for various nSCD values of IDPs. The solid lines correspond to the  DDx4 (solid) while the dashed lines are for the LAF1 at $\Gamma_e=3$. }
\label{Fig:idp}
\end{figure}

To model these IDP variants, we employ a modified Lennard–Jones (LJ) potential for non-bonded monomer interactions. This formulation allows the attractive interactions between $i$ and $j$ residues to vary independently from the short-range repulsion, scaled by their average hydropathy $\lambda = (\lambda_i + \lambda_j)/2$\cite{ashbaugh2008natively,weeks1971role}.

\begin{equation}
    U(r) = 
    \begin{cases} 
        U_{\text{LJ}}(r_{ij}) + (1 - \lambda) \varepsilon, & r_{ij} \leq 2^{1/6} \sigma \\ 
        \lambda U_{\text{LJ}}(r_{ij}), & \text{otherwise}
    \end{cases}
    \label{eq:modified_lj}
\end{equation}

where $U_{LJ}$ is
\begin{equation}
    U_{LJ} =  
  4 \epsilon_{LJ}\sum_{i>j}^{N} \left[\left(\frac{\sigma}{r_{ij}}\right)^{12} -\left(\frac{\sigma}{r_{ij}}\right)^{6} \right]
  \label{eq:lj}
\end{equation}

The hydropathy $\lambda$ values are based on the Urry scale \cite{urry1992hydrophobicity}, and the pair potential $U(r)$ is truncated and shifted to zero at a distance of $4\sigma$.

Figure~\ref{Fig:idp} presents the FEC for different variants of DDX4 (solid) and LAF1 (dashed). For sequences with lower nSCD values, the FEC shows a continuous transition from a coiled to a stretched conformation, whereas for sequences with higher nSCD, an abrupt transition is observed, which is similar to the trend seen in the PA  chain with longer block lengths, highlighting the connection between charge sequence patterning leading to an abrupt transition in the mechanical response.

\newpage

\begin{widetext}

\begin{table}[h]
    \caption{The table consists of the sequences of the LAF1 for a few nSCD values. Here, A, R, N, D, C, Q, E, G, H, I, L, K, M, F, P, S, T, W, Y, V are the short forms of the amino acids.}
    \label{Tab:laf}
    \begin{tabular}{|c|p{11cm}|}
        \hline
            \multicolumn{1}{|c|}{\textbf{LAF1}} & \multicolumn{1}{c|}{\textbf{Sequence}} \\
        
        \hline
        \multirow{3.5}{*}{nSCD = 0.003} &MGGSGGRGSGRNGDYYGSGPGYRAQDYGHGDSRRGGANSSGGNG \\
                                     &GNPGRAEGREGGGGANGYYSGGGANQASGGNRNGGADGGGGGDR \\
                                     &GGGNNGNRGDAGYREGNRRGDNNGNFGDNNSDRGLRGASDRDGN \\
                                     &RNYLGSRDRDVNGGGDYNYRNGGYRNRGDDGRSGDR \\
        \hline

        \multirow{3.5}{*}{nSCD = 0.131} &MGGDRAGYNNGDRLNRNGFRRLSGPNGAGGGNSDRRYYRNARRG \\
                                     &GGRGGRAGGYGYNGGRRRAQGSRGEGGGGGGSYDNGNRGEGSGR \\
                                     &GRNNDGGGDDNARNNDSHRDPGRVGNAGYGGGGGNGNGSYYGG \\
                                     &DNNDAGGSGSADGGGSNGDYDERGDSNGRDGNGQYDS \\
        \hline

        \multirow{3.5}{*}{nSCD = 0.263} &MDGGDDDGGQGGAGENLNDYYNGSSGGPGNSVAENSGGNGNDDG \\
                                     &GGDNRNNRYYSGGGDGRYGGGGQGANGGDDNAGDGYDGDNFGN \\
                                     &GNGRGYNNGSYGADASEGDRGSYRSSGRRRRRGSGGGGNYPGGGS \\
                                     &RRLNRNRGARGGAGGNRRRNARRDGNGYGGRRGRGH \\
        \hline

        \multirow{3.5}{*}{nSCD = 0.394} &MESNQSNNGGSGNAALNDGGDYVPPHLDGGDGGAAAAASAGGDE \\
                                     &DDGGAGGGGYDDGGGNSGGGGGGGYDEGYNDNDDDDNDGGSGG \\
                                     &YGDRRNYRRRGYNGGGGGGGNRGYNNNRGGGGGGYNRQRRGRG \\
                                     &GSSNFSRGGYNNRRRGSRNRGSGRSYNNRRRRNGGRGR \\
        \hline
    \end{tabular}
\end{table}

\begin{table}[h!]
    \caption{The table consists of the sequences of the DDX4 for a few nSCD values. Here, A, R, N, D, C, Q, E, G, H, I, L, K, M, F, P, S, T, W, Y, V are the short forms of the amino acids.}
    \label{Tab:laf}
    \begin{tabular}{|c|p{11cm}|}
        \hline
            \multicolumn{1}{|c|}{\textbf{DDX4}} & \multicolumn{1}{c|}{\textbf{Sequence}} \\
        
        \hline
        \multirow{5.5}{*}{nSCD = 0.002} &MDPTESFCANGGGAGSNKMGDGGNRATSDGLKNVFSMSNEADGRS \\
                                     &SDRSGNSNSGDPKDKGGETSRESRNGSGGFERGREPGNSGRNIGSSTI \\
                                     &EERLSRISGSDGTPYLDGSYSRERVSADHGGNMFYCFPSDNKGGSGL \\
                                     &RPNANEFNRERTSQVMDRPNGEGFFRSFYRHPVGGMESTYWGQSGS \\
                                     &GPERNGFGSDDGWRLSFRKDEWRCSDKDNGFNEFRKSGNGGERCT \\
                                     &GEGDF \\
        \hline

        \multirow{5.5}{*}{nSCD = 0.191} &MRNESEGEYCCEADDFVGNPSDNSYTRASTGSDDDRGFYFSFEAGG \\
                                     &SYDEFSSEGDEWDNNGSGGGDNFNESDFRGDNETGDPERGDNDGIE \\
                                     &KHGDPGISPLGRSPGMSRTTDTRGSMLGGSFEVGSGFLFTGESGSAS \\
                                     &MGWRGNSRFKFCRDMLQASKRINNNKVSGGRRSGESWCSGGREGT \\
                                     &GSSMRGRGPKFEGNRGRSPNNKAQGKRRRPFGGKHNRYNSERNGSL \\
                                     &NVSSGP \\
        \hline

        \multirow{5.5}{*}{nSCD = 0.284} &MNSEEYCGCEADDDFVGNPSNSYTRASTGSRDDEDGFFSYDEEEEEE \\
                                     &EFSSGDWDNNGSGGGEEDNFNSDFIKHRGDNETGDGSPGFGNGYFS \\
                                     &GDDPGISPLGSPGMSERRRRTTDDDTGGGSFVAGGLFTGSGSASMG \\
                                     &WSMFKRRFLCRRRKMLQASGNSINVSSNNKVSGGSGESWCSGGRRG \\
                                     &TGSSMKGGPFEGNGSPNNKAQGKRRRRRPFGGKHNRRRRRYNSEN \\
                                     &GSLNGP \\
        \hline

        \multirow{5.5}{*}{nSCD = 0.478} &MNESEEGYCACEDDEFVGNPSDNSEEYTASTGSDDDEEEEEEDEGFY \\
                                     &FSFAGGSYFSSGDWDNNGSGGGDNFNSDFGDNETGDPEGDNDGIEK \\
                                     &HGDDPGISPLGSPGMSRTTTRRGSMLGGSFEVGSGFLFTGSGSASMG \\
                                     &WRRRRGNSFKFGSWCSCDMLQASINNVGGNKNKVSSLRRRRRSGG \\
                                     &GTGSSMGKRRGPFGGNSPNNKAQGRRRRRRRRRPFGGKHNYNSRR \\
                                     &NGSSGP \\
        \hline
    \end{tabular}
\end{table}
\end{widetext}

\newpage


\renewcommand{\thesection}{\arabic{section}}






\providecommand{\latin}[1]{#1}
\makeatletter
\providecommand{\doi}
  {\begingroup\let\do\@makeother\dospecials
  \catcode`\{=1 \catcode`\}=2 \doi@aux}
\providecommand{\doi@aux}[1]{\endgroup\texttt{#1}}
\makeatother
\providecommand*\mcitethebibliography{\thebibliography}
\csname @ifundefined\endcsname{endmcitethebibliography}  {\let\endmcitethebibliography\endthebibliography}{}

\end{document}